\documentclass[aps,prb,reprint,groupedaddress]{revtex4-1}
\usepackage{graphicx}

\usepackage{color}
\usepackage{amsmath}
\usepackage{enumitem}
\usepackage{graphicx}
\usepackage[caption=false]{subfig}
\newcommand{\be}{\begin{equation}}
\newcommand{\ee}{\end{equation}}

\newcommand{\bea}{\begin{eqnarray}}
\newcommand{\eea}{\end{eqnarray}}

\newcommand{\la}{\left\langle}
\newcommand{\ra}{\right\rangle}

\newcommand{\lp}{\left(}
\newcommand{\rp}{\right)}
\newcommand{\sgn}{{\rm sgn\,}}

\renewcommand{\Im}{{\rm \, Im\,}}

\renewcommand{\vec}[1]{{\bf #1}}
\newcommand{\bra}[1]{\left\langle #1 \right|}
\newcommand{\ket}[1]{\left|#1\right\rangle}

\begin{document}

%\preprint{}

\title{Resonant electron-lattice cooling in graphene}


\author{Jian Feng Kong, Leonid Levitov}
\affiliation{Massachusetts Institute of Technology, Department of Physics, Cambridge, Massachusetts 02139, USA}
\author{Dorri Halbertal, Eli Zeldov}
\affiliation{Department of Condensed Matter Physics, Weizmann Institute of Science, Rehovot 76100, Israel}
%\email[Corresponding author: ]{abc@xyz}
%\thanks{}
%\affiliation{places} %Massachusetts Institute of Technology, Department of Physics, Cambridge, Massachusetts 02139}


\begin{abstract}
Controlling energy flows in solids through switchable electron-lattice cooling can grant access to a range of interesting and potentially useful energy transport phenomena. Here we discuss a tunable electron-lattice cooling mechanism arising in graphene due to phonon emission mediated by resonant scattering on defects in crystal lattice, which displays interesting analogy to the Purcell effect in optics. We argue that the dominant contribution to the electron-phonon cooling arises from hot carrier trapping on localized states at the defects. In contrast, phonon emission by a free electron, either near the defect or in pristine graphene, give subleading contributions. Resonant dependence of this process on carrier energy translates into gate-tunable cooling rates, exhibiting strong enhancement of cooling that occurs when the carrier energy is aligned with the electron resonance of the defect.
\end{abstract}

%\keywords{}

\maketitle


%\section{Introduction}

%electron-lattice cooling rate 
In 1946 Purcell discovered that %It is well known that 
bringing the energies of atoms in alignment with resonances in optical cavities can dramatically enhance the rate of spontaneous emission \cite{purcell_1946}. One way of understanding the enhancement is provided by Fermi's Golden Rule that mandates that the transition rate is proportional to the density of final states. The latter is enhanced in a cavity at resonance compared to a free-space density of states, providing means for controlling the light-matter coupling \cite{haroche_1989,yablonovitch_1987}.  Here we discuss an electron-phonon analog of Purcell effect: resonant enhancement of electron-lattice cooling occurring when carrier energies align with electron resonances at defects. Because of Purcell-type enhancement of the density of electronic states at the defects, the on-resonance electrons can emit phonons more efficiently, enhancing the electron-lattice cooling rate and making it gate-tunable. Furthermore, resonant scattering opens up an additional cooling pathway due to the possibility of carrier trapping on localized defects. The latter process, as we will show, boosts phonon emission and electron-lattice cooling.

\begin{figure}
	\subfloat[]{
		\includegraphics[width=0.5\linewidth]{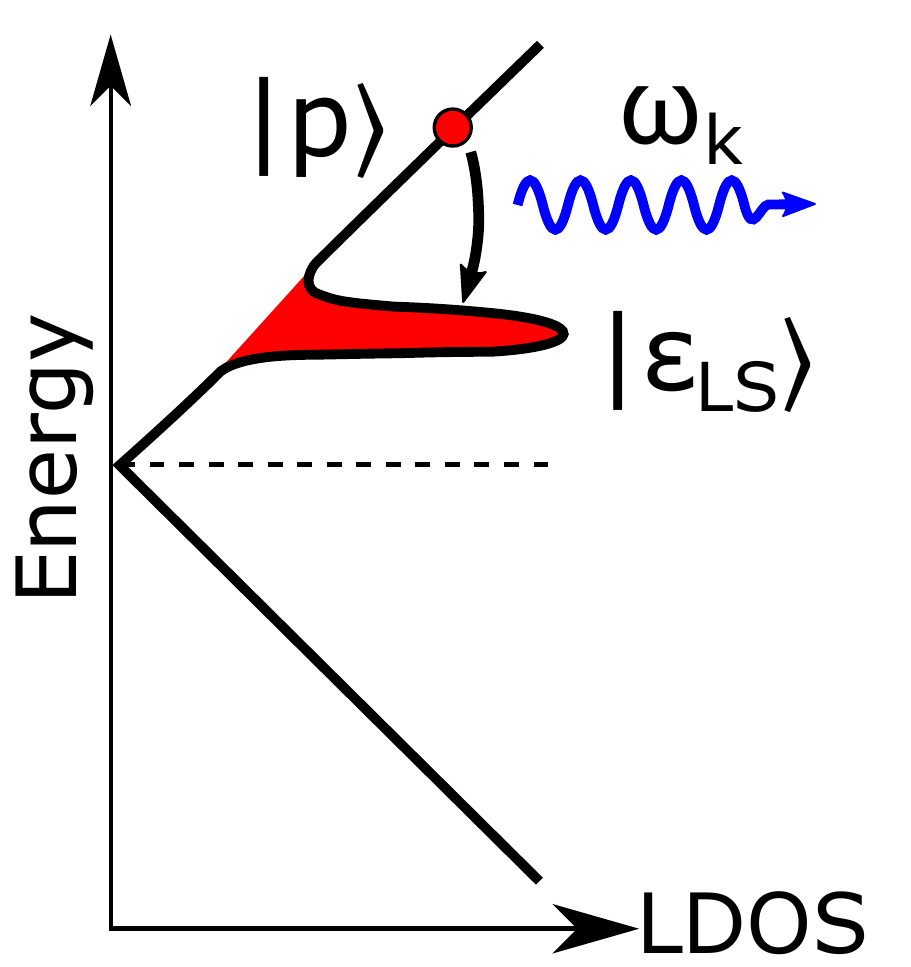}
	}
	\subfloat[]{
		\includegraphics[width=0.5\linewidth]{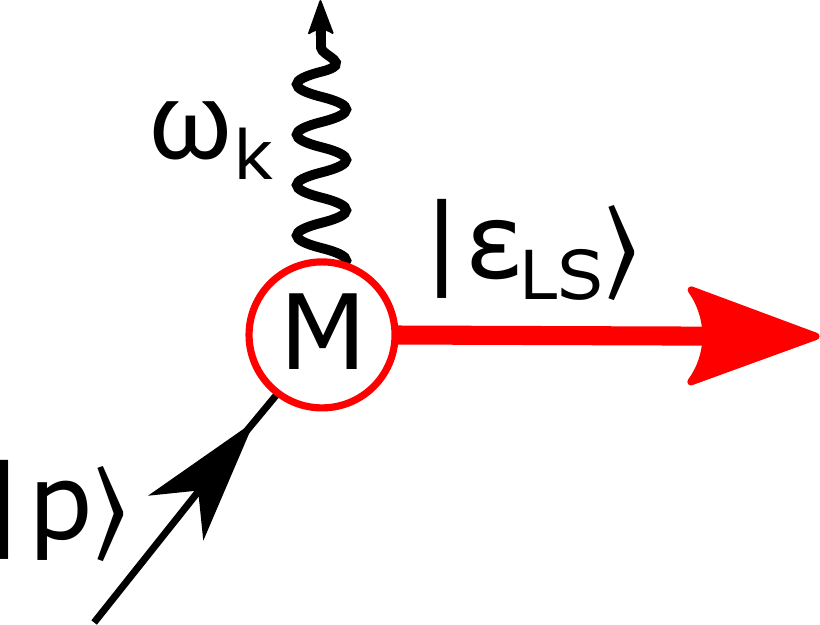}
		\label{fig:schematicMatEle}
	}
	\caption{(a) Schematics showing a hot electron of initial state $|\vec{p}\rangle$ being trapped by the impurity forming a resonant state $|\epsilon_{\text{LS}}\rangle$, emitting a phonon of energy $\hbar\omega_k$ in the process. (b) The same process shown as a Feynman diagram. The vertex represents the matrix element $M$ in Eq. (\ref{eqn:MatEle2}). Note that the outgoing state is the resonant state $|\epsilon_{\text{LS}}\rangle$.
	}
	\label{fig:scatproc}
	\vspace{-4mm}
\end{figure}

While these effects are completely generic, they become particularly important in graphene, a material in which energy relaxation pathways of nonequilibrium hot carriers are uniquely sensitive to minute amounts of disorder. In pristine graphene, electron-phonon scattering is suppressed and, as a result, the hot electron cooling is quite slow \cite{bistritzer_2009,tse_2009}. The introduction of defects completely changes the situation, giving rise to several different cooling mechanisms that can occur depending on the microscopic properties of the defects as well as system parameters such as carrier density and temperature. Resonant defects with energies near the Dirac point play a special role as the electronic density of Dirac states is low at these energies. As a result, carrier trapping on the defects strongly impacts cooling and phonon emission, as illustrated in Fig. \ref{fig:scatproc}.

Previous works on disorder-assisted electron-phonon scattering considered so-called ``supercollisions'' in which the excess recoil momentum is being absorbed by the impurity whereas the energy is carried away by a thermal phonon\cite{song_2015,song_2012,graham_2012,betz_2012,alencar_2014,tikhonov_2014}. This process, through phase-space-enhancement of electron-phonon scattering, gives rise to disorder assisted cooling. However,as this paper shows, it is not the only disorder-related cooling pathway. As we will see, resonant defects provide a fundamentally different cooling mechanism---phonon emission by an electron trapped by a defect--- that is distinct from the enhancement of phonon phase space through momentum-nonconserving scattering. 

Graphene-based nanoscale thermoelectric devices are of wide interest due to the unique electrical and thermal properties of this material\cite{pop_2012,xu_2016}. This work adds on to this exciting field by providing a new and controllable mechanism of hot carrier cooling. There are two main approaches to low-dimensional nanoscale thermal engineering: phononics engineering \cite{balandin_2012} and hot carriers manipulation. In the field of graphene phononics, the ideas such as gate tunability cooling \cite{shafranjuk_2016} and defect engineering have been investigated\cite{haskins_2011,anno_2017}. Complementary to that, our work explores the mechanism of hot carrier resonant cooling by localized defects. New cooling pathways in nano-devices with on-demand spatial dependence  through precision defect engineering are made possible by this new physical framework.

\begin{figure}
\subfloat[]{
	\includegraphics[width=0.95\linewidth]{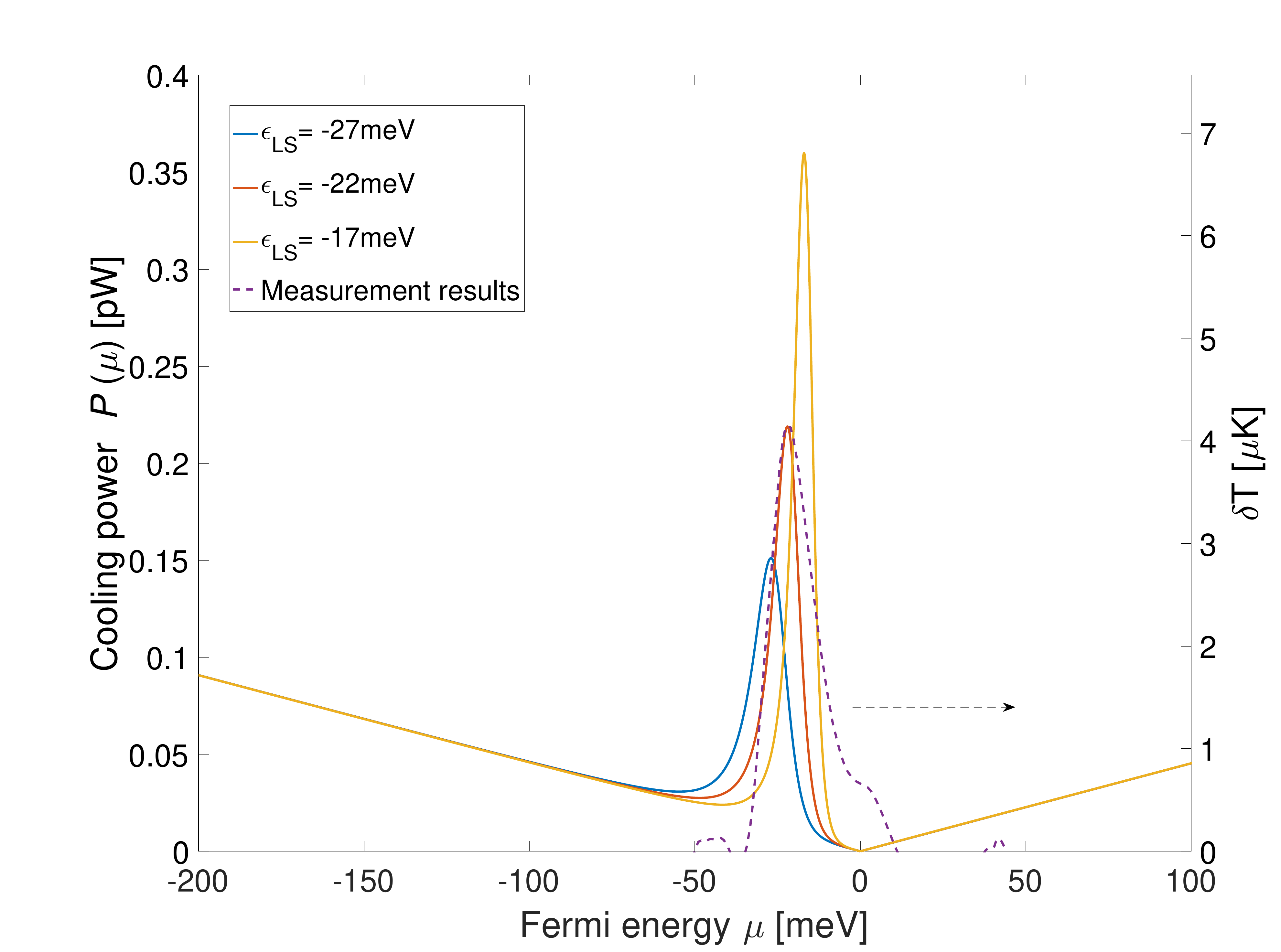}
	\label{fig:PepPartA}
}\\
\subfloat[]{
	\includegraphics[width=0.95\linewidth]{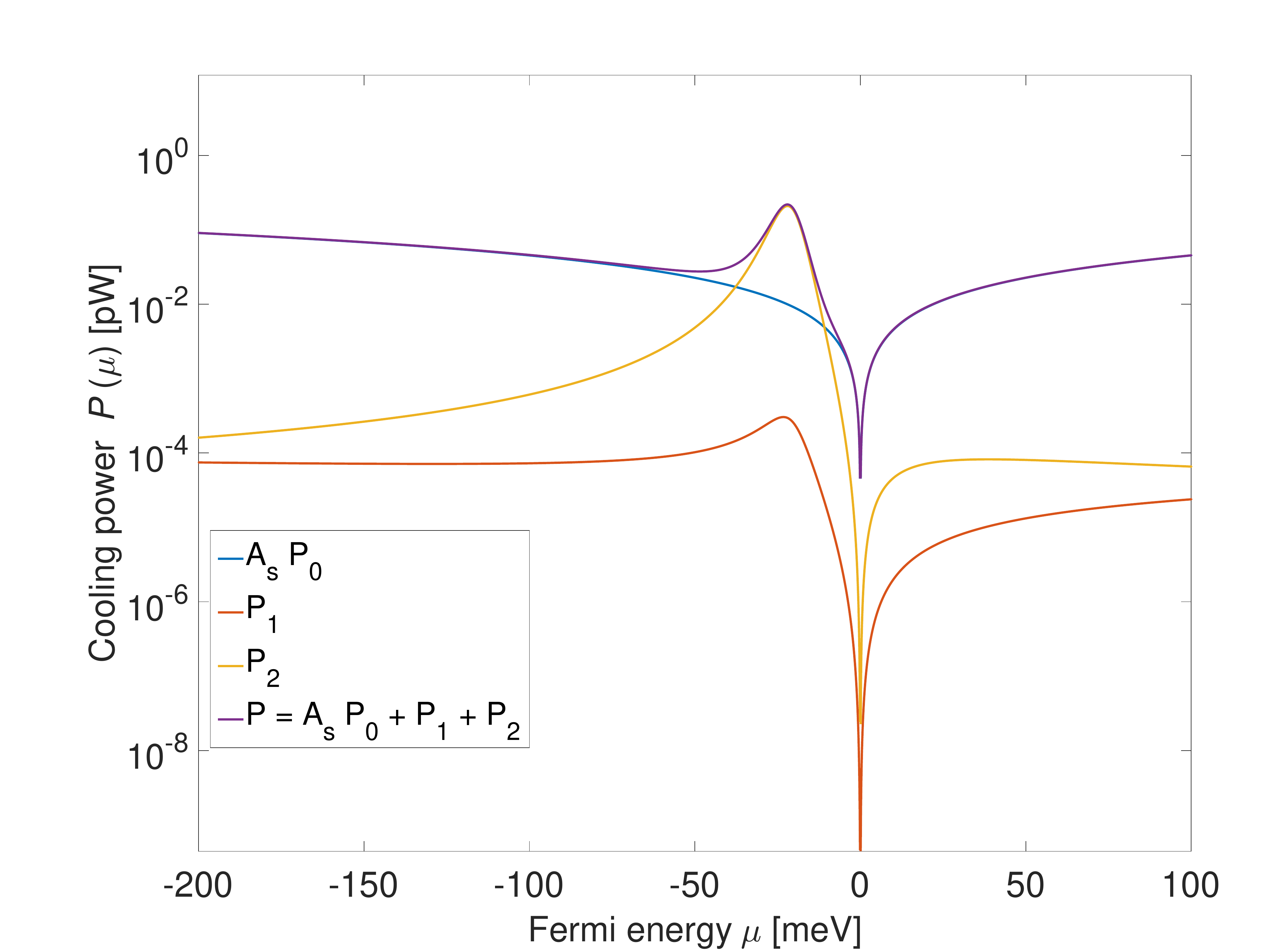}
	\label{fig:PeplogPlot}
}
	\caption{ A plot of the cooling power, in the presence of resonant scatterers vs. Fermi energy for several values of the scatterer %resonance 
	energy $\epsilon_{\text{LS}}$ [Eqs.\eqref{eqn:Pt},\eqref{eqn:Pep_4}]. Local temperature change [Eqn.(\ref{eqn:localT})], which is proportional to the cooling power, is shown on the right axies. The curves %scatterer contribution 
	are sharply peaked on resonance, falling off rapidly away from resonance.
%	The cooling power $P_{t}$ (Eq. \ref{eqn:Pt}) vs. Fermi energy for several values of $\epsilon_{\text{LS}}$ (Eq. \ref{eqn:Pep_4}). The contribution from $P_{ep}$ peaks near the energy of the resonant state and falls of as $1/\epsilon^2$ away from the resonance. 
	Purple dashed line shows experimental curve from \cite{halbertal_2017}, where peaks in cooling power due to resonant scatterers were observed near $\epsilon\approx -22$ meV. The intrinsic contribution %from momentum-conserving electron-phonon scattering 
	(Eq. (\ref{eqn:P0_2})) vanishes at the Dirac point, and remains low, compared to the peaks, throughout the range of Fermi energies plotted. %, becoming significant only beyond $|\epsilon|\sim 150$ meV. 
	%Inset illustrates the resonant cooling process due to phonon emission mediated by a resonant scatterer. Shown is the peak in the local density of states of the scatterer. %Localized disorder creates a peak in the local density of states near the Fermi energy. 
	%Hot electrons (red) get trapped in a quasi-bound state, from which a phonon is emitted resulting in resonantly enhanced electron cooling.
	(b) Semi-log plot showing the relative contributions of $P_0$, $P_1$ and $P_2$ for $\epsilon_{\text{LS}}=-22$ meV. $P_1$ is small throughout the range of $\mu$.
	}
	\label{fig:Pep_plot}
	\vspace{-4mm}
\end{figure}

Recently impurity-assisted electron-lattice cooling in graphene was imaged using nanoscale thermometry scanning probe technique \cite{halbertal_2016,halbertal_2017}. It was found that the dominant contribution to cooling arises from resonant scatterers with the energies of the resonances positioned near the Dirac point. It was conjectured that resonant scatterers  mediate phonon emission and cooling through the process of trapping band carrier in a localized state (LS) as illustrated in Fig. \ref{fig:scatproc}. While this picture seems plausible, the study reported in Refs.\cite{halbertal_2016,halbertal_2017} left a number of key questions unanswered, in particular the origin of the resonances and the extent to which resonant scattering can enhance the cooling rates. 
Below we present a microscopic picture of cooling due to phonon emission mediated by resonant scattering %discuss %estimate % analyze %calculate 
and estimate the cooling rate by evaluating 
the electronic cooling cross-section. % due to the presence of these resonant impurities 
The cooling rate is sharply enhanced when the Fermi energy is close to the resonance energy of one of the scatterers, turning off quickly when the Fermi level is detuned from the resonance energy. Disorder-assisted resonant cooling is found to dominate over the intrinsic contribution due to momentum-conserving electron-phonon processes. This, along with the ON/OFF switching behavior seen near the resonance, presents the novel possibility of gate-tunable cooling.

\section{Model of electronic cooling}
Graphene is known to host a wide variety of atomic-scale defects which can act as resonant scatterers, which can trap electrons in quasibound states \cite{pereira_2006,shytov_2009,ni_2010,titov_2010}. \emph{Ab initio} and STM studies \cite{wehling_2009,brihuega_2016,mao_2016} have shown
that quasi-bound states with energies near the Dirac point arise in a robust manner when adatoms or polar groups like H, F, ${\rm CH_3}$ or OH bind
covalently to carbon atoms, transforming the trigonal \emph{sp}$^2$ orbital to the tetrahedral \emph{sp}$^3$
orbital. Each transformed C atom gives rise to a vacancy in the $\pi$-band, producing a
quasibound state localized near the defect. The energy of such a localized state depends on the adatom type, taking values $\epsilon_{\text{LS}}\sim 10-100\,{\rm meV}$ i.e., positioned in the direct vicinity of the Dirac point\cite{wehling_2009,brihuega_2016,mao_2016}. In transport, such defects act as resonant scatterers, with the scattering cross section exhibiting a sharp resonance at $\epsilon=\epsilon_{\text{LS}}$. %positioned near the Dirac point.  
In contrast, the defects having other symmetries (e.g. adatoms positioned between two C atoms or at a hexagon center) typically form resonances far away from the Dirac point.

Here we shall consider phonon emission by carriers in the presence of such resonant scatterers, %as these are responsible for electronic cooling, 
assuming that the lattice is at a constant temperature of $T_{p}$, forming a thermal phonon bath. We show that the resonance in the local density of states gives rise to enhanced phonon emission in a manner similar to how spontaneous photon emission is enhanced by the optical cavity resonances in the Purcell effect. The Hamiltonian is given by a sum of the electron and phonon parts, and the electron-phonon interaction,  $H = H_{\rm el} + H_{\rm ph} + H_{\text{el-ph}}$, where
\be
H_{\text{ph}}=\sum_\vec{k}\omega_\vec{k}b_\vec{k}^\dagger b_\vec{k} %\sum_{\vec{k}}\ a_\vec{k}^\dagger\; \sigma\cdot \vec{q} \;a_\vec{k}
,\quad
H_{\text{el-ph}} = \sum_{\vec{k},\vec p} g\sqrt{\omega_\vec{k}}b_\vec{k}a^\dagger_{\vec{p}+\vec{k}}a_\vec{p} + \text{h=H.c.}
\ee
and the electron Hamiltonian $H_{\text{el}}$, describing free carriers and their interaction with the defects, is discussed below. 
%\bea
%H &=& H_{el} + H_{ph} + H_{el-ph} + H_{dis} \\
%&=& v_F\sum_{\vec{k}}\ a_\vec{k}^\dagger\; \sigma\cdot \vec{q} \;a_\vec{k} + \sum_\vec{q}\omega_\vec{q}b_\vec{q}^\dagger b_\vec{q} + H_{el-ph} + H_{dis}
%\eea
%where
%\be
%H_{el-ph} = \sum_\vec{q} g\sqrt{\omega_\vec{q}}b_\vec{q}a^\dagger_{\vec{p}+\vec{q}}a_\vec{p} + h.c.
%\ee
%is the electron-phonon coupling, and we consider the disorder potential to be due to a single impurity localized at the origin without loss of generality:
%\be
%H_{dis}= V\sum_{\vec{p},\vec{p'}} a_\vec{p}^\dagger a_\vec{p'}
%\ee
From the above Hamiltonian, we can calculate the energy dissipation rate as %which can be written as:
\bea \nonumber 
&&
P = \sum_{\vec p, \vec{p'},\vec{k}} \omega_{\vec{k}} W_{\vec{p}, \vec{p'} \vec{k}} (1-n_{\vec{p}'})n_{\vec{p}}(N_k+1) \delta(\epsilon_{\vec{p}'}+\omega_{\vec{k}}-\epsilon_{\vec{p}}) 
\\ \label{eqn:CPS}
&&-\sum_{\vec p,\vec{p'},\vec{k}} \omega_{\vec{k}} W_{\vec{p'}\vec{k}, \vec{p}} (1-n_{\vec{p}})n_{\vec{p}'} N_k \delta(\epsilon_{\vec{p}'}+\omega_{\vec{k}}-\epsilon_{\vec{p}})
\eea
where $n_{\vec{p}}$, $n_{\vec{p}'}$ and $N_k$ are Fermi and Bose distributions for electrons and phonons with momenta $\vec{p}$, $\vec{p'}$ and $\vec{k}$, and with energies $\epsilon_{\vec{p}}=\epsilon_p$, $\epsilon_{\vec{p}'}=\epsilon_{p'}$, and $\omega_{\vec{k}}=\omega_k$ respectively. The scattering cross-section in Eq. (\ref{eqn:CPS}) equals
\be
W_{\vec{p}, \vec{p'} \vec{k}} = W_{\vec{p'}\vec{k}, \vec{p}} =\frac{2\pi}{\hbar} |M_{\vec{p},\vec{p'}}|^2
\ee
where an incoming electron $\ket{\vec{p}}$ is scattered into the outgoing state $\ket{\vec{p'}}$, emitting or absorbing a phonon with momentum $\vec{k}=\vec{p}-\vec{p'}$. 
%$M_{\vec{p},\vec{p'}}$ is the matrix element for this process and we consider three processes that contribute to it. 

The matrix element $M_{\vec{p},\vec{p'}}$, describing phonon emission in the presence of a defect, can be written as a sum of three terms:
\be
\label{eqn:MatEle}
M_{\vec{p},\vec{p'}}=\bra{\vec{p'}}M(k) G_0t + tG_0M(k) + tG_0M(k)G_0t \ket{\vec{p}}
\ee
where $t$ is the T-matrix for the defect, 
evaluated at the energies of the in and out states as discussed below, $G_0$ is the bare electronic Green's function, and
\be
M(k)\equiv \bra{\vec{p}',\vec{k}}H_{\text{el-ph}}\ket{\vec{p}} = g \sqrt{\omega_k} \delta(\vec p-\vec p'-\vec k)
\ee
is the bare electron-phonon scattering matrix element. The three terms in Eq.\eqref{eqn:MatEle} correspond to resonant scattering at the defect before, after, and both before and after a phonon emission process. The overall process is illustrated in Fig. \ref{fig:schematicMatEle}.

We will find that %by tuning Fermi energy close to the energy of the resonant state, 
the contribution to the cooling rate due to resonant scattering, taken on-resonance,  is large compared to the contribution from bare momentum-conserving electron-phonon scattering in the absence of disorder. This comparison, which provides a justification for focusing on the disorder-assisted elecron-phonon scattering processes, will be made after the resonance-enhanced cooling rate is evaluated. %calculated.
%We note that the phonon emission process occurring in the absence %undressed by 
%of impurity scattering is neglected here as its contribution to the cooling rate is small compared to the impurity assisted process. \addLL{[please elaborate, this is an important point]}

The bare electron propagator is given by
\be
G_0(\vec{p},\epsilon) = \frac{1}{\epsilon-H_0(\vec{p})+i0}
\ee
with the free-particle tight-binding Hamiltonian given by a $2\times 2$ matrix in the $A/B$ sublattice pseudospin basis
\be
H_0=\lp\begin{array}{cc} 0 & t_h f(\vec p)\\ t_h f^*(\vec p) & 0\end{array}\rp
,\quad
f(\vec p)=\sum_{i=1}^3 e^{i\vec p\vec e_i}
.
% f(\vec k)=e^{i\vec k\vec e_1}+e^{i\vec k\vec e_2}+e^{i\vec k\vec e_3}
\ee
%In this representation, 
Here $\vec e_i$ are vectors connecting neighboring C atoms %with their neighbors 
and $t_h$ is the nearest-neighbor hopping parameter. 

\section{T-matrix for resonant defects}
To describe resonant scatterers %can be described in this framework by 
we introduce diagonal on-site disorder potential $V(\vec x)=\sum_iV\frac12 (1\pm\sigma_z)\delta(\vec x-\vec x_i)$ with the signs plus and minus corresponding to defects positioned on sites A or B, respectively. Resonant character of this disorder potential becomes prominent in the limit $V\gg W$, where $W\approx 6eV$ is the graphene bandwidth. This is evident from 
 % $H_0(\vec{p})=v_F \sigma\cdot\vec{p}$. 
 the T-matrix, which describes the defect potential renormalized by multiple scattering processes. For a single defect, taken without loss of generality on an A site at $\vec x=0$, T-matrix equals
\be
t(\epsilon) = \frac{\tilde V}{1-\tilde V\sum_\vec{p} G_0(\epsilon,\vec{p})} 
= \frac{\pi v_F^2}{\epsilon\ln\frac{iW}{\epsilon}+\delta} \frac{1 +\sigma_z}2
\ee
where $\tilde V=\frac12(1 +\sigma_z)V$ 
%$W\approx 6eV$ is the bandwith for which the Dirac Hamiltonian is valid, 
and $\delta=\pi v_F^2/V \ll W$. The T-matrix has a resonance centered at $\epsilon_{\text{LS}} \approx -\delta/\ln(W/\delta)$, which corresponds to the energy of localized state. % (LS), which 
The energy of the resonance $\epsilon_{\text{LS}}$ is small, %close to the Dirac point, 
with $\delta$ parameterizing the detuning from the Dirac point.

When the defect concentration is low and defect potential $V$ is large compared to $W$, each defect hosts a single resonance state with energy $\epsilon_{\text{LS}}$ close to the Dirac point, broadened due to hybridization with the states in the Dirac continuum. For a strong defect potential $V\gg W$ the energy $\epsilon_{\text{LS}}$ is much smaller than the bandwidth $W$ and it has an opposite sign to that of $V$. This resonance has a half-width of $\gamma\approx \pi \delta/(2\ln(W/\delta))$, and is thus small when $\delta$ is small.

The contribution of %an impurity (or 
defect to the single-particle density of states is given by
\be
\frac{1}{\pi} \Im t(\epsilon) = \frac{\pi v_F^2 |\epsilon|/2}{(\epsilon\ln(W/|\epsilon|)+\delta)^2+(\pi\epsilon/2)^2}
\ee
where we used the identity $\ln(iW/\epsilon) = \ln(W/|\epsilon|) + i\pi \sgn(\epsilon)/2$. This expression can be viewed as the density of states of pristine graphene $\sim |\epsilon|$ modulated by a resonant energy dependence due to the defect. The peak in the energy dependence of $\Im t(\epsilon)$, positioned near the Dirac point, corresponds to the defect resonance state.

%\subsection{Spatial Dependence of Resonance States}
%
%We can get a understanding of the spatial dependence of such resonance states by considering the following low-energy Dirac Hamiltonian with an impurity potential
%\be
%H = \begin{bmatrix}
%	0 & -iv_F \partial_+ \\
%	-iv_F \partial_- & V \delta(\vec{x})
%\end{bmatrix}
%\ee
%where $\partial_\pm = \partial_x \pm i \partial_y$, and we place the impurity at the origin without loss of generality. For a large on-site $V$, the Hamiltonian hosts a single zero-energy eigenstate of the form
%\be
%\psi(x,y) = \frac{a}{x-iy} \begin{pmatrix}
%	1 \\
%	0
%\end{pmatrix}
%\ee
%and the particle-hole symmetry becomes exact in the large $V$ limit. The wavefunction $\psi(x,y)$ has a power law tail and a log-diverging normalization, and is thus marginally delocalized. The wavefunction has non-vanishing values on the opposite sublattice to that of which the defect is residing, as one would intuitively expect. Also note that this zero-energy state survives when many defects are placed on the same sublattice, since the wavefunction resides on the opposite sublattice.

%\section{Cooling Power Spectrum}

%In this section we proceed to calculate the cooling power spectrum based on the model set up in the previous section. 
Next we proceed to calculate the cooling power. 
It will be seen that in realistic regimes the cooling power dependence on electron Fermi energy shows a peak that mimics the defect density of states, with a prefactor that depends on the electron and phonon energy distributions. Microscopically, there are two separate resonant processes. In one, an electron emits phonon after being trapped on the localized state at a defect. Namely, a resonant scatterer traps band electrons on the quasibound state, and the energy difference is released to phonons. In another process, a freely moving electron emits a phonon before or after being scattering by a resonant defect. In this case, due to breaking of translation symmetry by the presence of defects, momentum does not have to be conserved as these defects can absorb recoil momentum from the phonons. This effect boosts the available phase space for the outgoing states, making the cooling rate higher than in pristine graphene, as described in Song \emph{et. al.} \cite{song_2012} for weak disorder potential. As we will see, in our case the latter effects provides a relatively small contribution to the cooling power in comparison to resonant cooling through the processes involving electron trapping on the defect. 
%, and we find that different regimes lead to different $T$ dependence of $P(\epsilon)$.

%\subsection{Low $T$ and $\epsilon_F$}
%
%In the following analysis we will 

\section{Evaluation of matrix elements}
To evaluate the cooling power, we need to first evaluate the matrix element in Eq. \eqref{eqn:MatEle}. We will focus on the experimentally relevant regime of electron and phonon temperatures small compared to the resonance energy $\epsilon_{\text{LS}}$ and width $\gamma=\pi\delta/(2\ln(W/\delta))$. In this case, since the change of electron energy
%, which is of order $k_B T$ and $\epsilon_F$, whichever is larger, 
is small compared to  $\epsilon_{\text{LS}}$ and $\gamma$, the process is quasi-elastic.
%We first consider the regime where $T$ and Fermi energy $\epsilon_F$ are much smaller than the resonance width $\gamma=\pi\delta/(2\ln(W/\delta))$. In this regime, the change of electron energy, which is of order $k_B T$ and $\epsilon_F$, whichever is larger, is small compared to $\gamma$ and thus the process is quasi-elastic. 
Also, at not too low temperatures the phonon momentum values $\vec{k}$ are typically large compared to electron in and out momentum values $\vec{p}$ and $\vec{p'}$, which are of order $k_F$. This allows us to approximate $G_0(\epsilon,p)\sim \pm 1/\sigma v k$. The first two terms of Eq. \ref{eqn:MatEle} then combine to give a commutator
\be
- M(k) t(\epsilon) \frac{[\vec{\sigma}\cdot \vec{k}, \sigma_3]}{2 k^2} = M(k) t(\epsilon) \frac{i \vec{\sigma} \times \vec{k}}{v k^2}
\ee
Here $M(k)=g\sqrt{\omega_k}$ is the bare electron-phonon interaction matrix element.

The third term in Eq. \ref{eqn:MatEle} can be evaluated by integrating the product of two Greens
functions over internal electron momenta $k<q<k_0=W/v_F$, giving
\be
\begin{split}
	&tG_0M_0G_0t \approx \frac{M(k)}{2\pi v_F^2}  \frac{1+\sigma_3}{2} t(\epsilon') t(\epsilon) \int_{k}^{k_0} \frac{dq}{q} \\
	& = \frac{M(k)}{2\pi v_F^2}  \ln\frac{W}{v_F k}\frac{1+\sigma_3}{2} t(\epsilon') t(\epsilon)
\end{split}
\ee
Summing the three terms in Eq.\eqref{eqn:MatEle} gives %now becomes
\begin{equation}
\label{eqn:MatEle2}
%\begin{split}
M_{\vec p,\vec p'} = M(k) \frac{i\sigma\times \vec k}{v_F k^2}t(\epsilon) \\
 +\frac{M(k)}{2\pi v_F^2}  %\ln\frac{W}{v_F k}
\ln\frac{k_0}{k} \frac{1+\sigma_3}{2} t(\epsilon') t(\epsilon)
 .
%\end{split}
\end{equation}
%where we suppressed the delta function $\delta(\vec p-\vec p'-\vec k)$. 
%\addLL{[there's no delta function]} 
%functions over internal electron momenta $k<q<W/v$, giving
%\be
%\begin{split}
%	&tG_0M_0G_0t \approx \frac{M(k)}{2\pi v_F^2}  \frac{1+\sigma_3}{2} t(\epsilon') t(\epsilon) \int_{k}^{W/v} \frac{\mathrm{dq}}{q} \\
%	& = \frac{M(k)}{2\pi v_F^2}  \ln\frac{W}{v_F k}\frac{1+\sigma_3}{2} t(\epsilon') t(\epsilon)
%\end{split}
%\ee
%Summing the three terms in Eq.\eqref{eqn:MatEle} gives %now becomes
%%
%\begin{equation}
%\label{eqn:MatEle2}
%%\begin{split}
%M_{\vec p,\vec p'} = M(k) \frac{i\sigma\times k}{v_F k^2}t(\epsilon) \\
% +\frac{M(k)}{2\pi v_F^2}  \ln\frac{W}{v_F k}\frac{1+\sigma_3}{2} t(\epsilon') t(\epsilon)
%%\end{split}
%\end{equation}
%where we suppressed the delta function $\delta(\vec p-\vec p'-\vec k)$. 
The two contributions in Eq.\eqref{eqn:MatEle2} can be compared directly by ignoring the matrix structure. Since the process is quasi-elastic, the difference between $\epsilon$ and $\epsilon'$ is inessential and the %since the process is quasi-elastic. At a first glance the 
second term, which represents resonant trapping, dominates over the first term.

To better understand the competition between the two terms in Eq. (\ref{eqn:MatEle2}), we consider their ratio 
% of the first and second term
% is of order
\be
\frac{M_1}{M_2}%\sim \frac{v_F}{t(\epsilon) k}
\approx \frac{2s}{v_F}\frac{\epsilon\ln\frac{iW}{\epsilon}+\delta}{ k_BT \ln\frac{k_0}{k}}
\ee
where $s$ is the acoustic sound velocity $s\approx 2\times 10^4\,{\rm m/s}$. %and we suppressed a log factor. 
Since the velocity ratio $s/v$ is quite small, %which is positioned near the Dirac point, 
the second term in Eq.\eqref{eqn:MatEle2} will indeed dominate for the energies near resonance, $\mu\sim\epsilon_{\text{LS}}$ and at not too low temperatures. However, a different behavior is expected for energies away from the resonance, since the second term fall off faster than the first term  ($1/\epsilon^2$ vs. $1/\epsilon$). As a result the first term can win at large enough $\epsilon$. For a crude estimate, taking the detuning %of $\epsilon$ 
from resonance on the order of $\epsilon_{\text{LS}}$ we see that the first term becomes relevant when 
%the resonance energy, $\epsilon\sim \epsilon_{\text{LS}}$ 
%he relative importance of the 
 $k_BT\gtrsim (s/v_F)\epsilon_{\text{LS}}$. Taking $\epsilon_{\text{LS}}\sim 30\,{\rm meV}$ gives %the condition 
 $T>5\,{\rm K}$, which is close to the measurement temperature in \cite{halbertal_2016,halbertal_2017}. This analysis indicates that the second term dominates at resonance, whereas the first term dominates away from resonance.
 %behavior of cooling power on resonance is described by the second term, however the behavior } %Specifically

%LL STOPPED HERE

%\addLL{[this is close to the temperature at which measurement is carried out so let's check how the log factors change the estimate]}

%For temperatures of interest the phonon momentum $k_p \approx k_B T_e/s$, where $s\approx 2\times 10^4 m/s$, the sound velocity, is much greater than Fermi momentum for electron Fermi energy on resonance with the defect, $\epsilon_{\text{LS}} \approx \epsilon_F$. We can then ignore the first term as compared to the second term. 

\section{Cooling power}
Next we show that resonant energy dependence of phonon emission translates into a resonant dependence of cooling rate as a function of carrier doping. 
After plugging the matrix element given in Eq.\eqref{eqn:MatEle2} into the expression for the cooling power, Eq.\eqref{eqn:CPS}, and averaging over $\vec p$ and $\vec p'$ angles, we find that the contributions of the first and second terms separate %without generating any 
whereas the cross terms vanish under trace. Therefore, in the regime of interest $T_{\mathrm{BG}}<T<\epsilon_{\mathrm{LS}}$, the two terms in Eq. (\ref{eqn:MatEle2}) give independent contributions to the cooling power. These contributions describe the two distinct processes discussed above. In the first case, phonon is emitted by a freely moving electron before or after resonant scattering. In the second case, phonon is emitted by an electron which is trapped on a defect.

To evaluate the second contribution, which, according to Eq. (13), is expected to dominate over the first contribution, we use the identity: $\sum_{\vec{p},\vec{p'}}=N\int {\rm d\epsilon}\, {\rm d\epsilon'}\, \nu(\epsilon) \nu(\epsilon')$, where $\nu(\epsilon)$ is the density of states per spin per valley and $N=4$ is the spin and valley degeneracy, assuming unit area. The energy conservation delta function can be used to evaluate one of the integrals, the other integral can be evaluated by using the quasi-elastic approximation: $\int_\infty^\infty {\rm d\epsilon}\, g(\epsilon)(n(\epsilon)-n(\epsilon-\omega))\approx -g(\mu)\omega$, where $g(\epsilon)$ is an arbitrary smooth function of $\epsilon$, and the identity $ n(\epsilon)(1-n(\epsilon-\omega))=N_\omega^e (n(\epsilon-\omega)-n(\epsilon))$, with $N_\omega^e$ being the Bose function evaluated at the electronic temperature. The final integral over $\omega_k$ is of the form
\be
\int_{0}^{\infty} {\rm d\omega}\, \omega^4 (N^e_\omega - N^p_\omega) = 24 \zeta(5) k_B^5 \left(T_e^5 - T_p^5\right)
\ee
where $N_\omega^p$ is the Bose function evaluated at the lattice temperature. Putting everything together,
%The second term gives the contribution
we arrive at
\begin{equation}
P_2(\mu) = A(\mu)k_B^5 \left( T_e^5 - T_p^5\right)
\label{eqn:Pep_3}
\end{equation}
for the cooling power per defect. 
Here $\mu$ is the Fermi energy and
\be
A(\mu) = \frac{48\zeta(5)}{\pi^2}\frac{D^2}{\hbar^3 \rho s^4}\frac{(\pi/2)^4 \mu^2 \ln^2\frac{k_0}{k_p}}{[(\mu \ln\frac{W}{|\mu|}+\delta)^2 + \frac{\pi^2}4 \mu^2]^2} \,,
\ee
where $k_p \approx k_B T_e/s$ is the typical momentum of emitted phonon,
and we used the relation between the electron-phonon coupling constant and graphene deformation potential $g^2=D^2/2\rho s^2$ with $D\approx 50 eV$, and $\rho$ is the mass density of the graphene sheet. % and %correspondingly 
The quantity $P_2(\mu)$ vanishes in equilibrium as a result of detailed balance, but is non-zero when the system is driven out of equilibrium, as one would expect. For a numerical estimate we use the value $D^2/(\hbar^2\rho s^2 v_F^2)=1.86\times 10^{19} J^{-1}$. After scaling Fermi energy by $1$ meV and temperature by $1$ K, we evaluate the numerical factors to obtain
\be
P_2(\mu)=\frac{\mu^2 \ln^2\frac{k_0}{k_p}(T_e^5-T_p^5){\rm [1 meV]^2}}{[(\mu\ln\frac{W}{|\mu|}+\delta)^2+\frac{\pi^2}4\mu^2]^2{\rm [1 K]^5}} \times 260\, {\rm  fW}
.
\label{eqn:Pep_4}
\ee
%\addLL{how is this result derived?}
%for the cooling power per defect, where $k_p \approx k_B T_e/s$ is the typical momentum of emitted phonon. % momentum, 
%$s\approx2\times10^4$ m/s is the graphene sound velocity.
This contribution peaks near the resonance energy, falling off as $1/\mu^2$ at large detuning. % $\mu\gg\gamma$. 

The contribution of the first term in Eq.\eqref{eqn:MatEle2} can be evaluated in a similar manner, giving\cite{song_2015} 
\be
P_1(\mu)=9.62\frac{g^2 k_B^3\lp T_e^3-T_p^3\rp}{2 \hbar^3 v_F^2}\nu^2(\mu) |t(\mu)|^2
\ee
where $\nu(\mu)=|\mu|/2\pi \hbar^2 v_F^2$ is the density of states. After scaling Fermi energy by $1$ meV and temperature by $1$ K this expression becomes
\be
P_1(\mu)= \frac{\mu^2 (T_e^3-T_p^3) [1 {\rm K^{-3}}]}{(\mu \ln\frac{W}{|\mu|}+\delta)^2 + \frac{\pi^2}4\mu^2} \times 0.28\,{\rm  fW} \,.
\ee
The total cooling rate per defect is then given by $P_{\text{tot}}(\mu)=P_1(\mu)+P_2(\mu)$. 

%\addLL{
It is instructive to compare the cooling rates due to resonant scattering with the intrinsic contribution of % momentum-conserving cooling rate expected for 
pristine graphene \cite{bistritzer_2009,tse_2009}. The cooling power per unit area due to momentum-conserving processes can be written as
\be
P_0(\mu)=N\nu^2\sum_{\theta_\vec p,\theta_\vec p'}\frac{2\pi}{\hbar} g^2\omega_k %\frac{1+\cos\theta_{\vec p,\vec p'}}2
|\la \vec p'|\vec p\ra|^2 (N^e_{\omega_k}-N^p_{\omega_k})\omega_k^2
%|M(\vec p-\vec p')|^2
\ee
with the phonon energy $\omega_k=\hbar s|\vec p-\vec p'|$ %=s2k_F|\sin(\theta_{\vec p,\vec p'}/2)|$ 
and the coherence factor $|\la \vec p'|\vec p\ra|^2=\cos^2 (\theta_{\vec p,\vec p'}/2)$. Here $\sum_{\theta_\vec p,\theta_\vec p'}$ denotes averaging over the Fermi surface through $\oint\oint\frac{d\theta_\vec p d\theta_\vec p'}{(2\pi)^2}$. We parametrize 
\be
|\vec p-\vec p'|=2k_F x,\quad
0<x<1
\ee 
Writing $|\la \vec p'|\vec p\ra|^2=1-x^2$ and $d\theta_{\vec p,\vec p'}=\frac{2dx}{\sqrt{1-x^2}}$ we can express the cooling power as
\be
P_0(\mu)=N\nu^2\frac2{\pi}\int_0^1 dx \sqrt{1-x^2}\frac{2\pi}{\hbar} g^2(N^e_{\omega_k}-N^p_{\omega_k})\omega_k^3
,
\ee
where $\omega_k=2sxk_F$. 
%}

%\addLL{
This expression behaves differently dependening on whether the temperature $T$ is greater or smaller than $T_{\rm BG}=\hbar sk_F$. For $T\gg  T_{\rm BG}$ we can approximate the Bose distribution as $N_{\omega_k}\approx \frac{T}{\omega_k}$. Using the identity $\int_0^1 dx x^2 \sqrt{1-x^2}=\frac14\Gamma^2(3/2)=\frac{\pi}{16}$, we obtain %which gives
%The cooling power per unit area due to momentum-conserving processes is given by 
\be
P_0(\mu)=B(\mu) k_B(T_e-T_p),
\ee 
with $B(\mu) = \pi N\hbar g^2 \nu^2(\mu) k_F^2 s^2$. The numerical factors can be evaluated to give
\be
P_0(\mu) = 9.2 \times 10^{-3} \frac{(T_e-T_p)%\,[{\rm K}^{-1}] \, 
\mu^4}{[{\rm 1K}]  [{\rm 1meV}]^4}\, {\rm fW \, \mu m}^{-2} \,.
\label{eqn_P0}
\ee
In the limit $T\ll T_{\rm BG}$ the integral over $x$, which is dominated by $x\ll 1$, can be estimated as %}

\be
\begin{split}
 P_0(\mu)&=N\nu^2\frac2{\pi}\int_0^\infty dx \frac{2\pi}{\hbar} g^2(N^e_{\omega_k}-N^p_{\omega_k})\omega_k^3
\\
&=4N\nu^2 g^2\frac{3\zeta(4)}{\hbar^2 sk_F}k_B^4(T_e^4-T_p^4) \\ 
&= 3.0 \times \frac{(T_e^4-T_p^4)|\mu|}{[1 \, {\rm K}^4][1\,{\rm meV}]} \, {\rm fW} \, \mu {\rm m}^{-2} \,.
\label{eqn:P0_2}
\end{split}
\ee

To compare $P_0(\mu)$ to the resonant scattering contribution, Eq.\eqref{eqn:Pep_4}, we have to consider the physical measurement process. In such a measurement, the tip picks up the thermal signal from a sensing region with area $A_s$, and thus the cooling power contribution from $P_0$ is given by $A_s P_0$.  With the experimentally realistic value of $A_s=100\times100\,{\rm nm}^2$, and assuming that there is only 1 defect in such a region, we construct the quantity 
\be
P(\mu) = A_s P_0(\mu) + P_1(\mu) + P_2(\mu) .
\label{eqn:Pt}
\ee

To accurately depict the behavior of this intrinsic contribution at large Fermi energies, where $T_{\mathrm{BG}}$ is greater than $T$, we will use Eq. (25) in the ensuing discussion. (Both Eq. (25) and Eq. (24) give contributions to the cooling rate which are insignificant near the LS resonance.) We plot the cooling power Eq. (\ref{eqn:Pt}) in Fig. \ref{fig:PepPartA}, for three different values of $\epsilon_{\text{LS}}$, and with the parameters $\ln (k_0/k_p) = 5.6$, $T_e=4.25$ K, $T_p=4.2$ K. The sharp peaks near $\epsilon_{\text{LS}}$ come from the contribution of $P_2$ while $P_1$ affects the tails of the peak, as can be seen in Fig. \ref{fig:PeplogPlot}. The intrinsic contribution $P_0$ is smaller than that of the defects at all energies shown in Fig. \ref{fig:PepPartA}. As a result, the resonant peak in the cooling rate due to the defects dominates near charge neutrality.

\section{Comparison to experiment}
To make a direct comparison with the measurement results reported in Ref. \cite{halbertal_2017} which employ a nanoscale probe to detect local changes in temperature, we convert the cooling power to local temperature change as follows. We assume that the power dissipated at the defect generates phonons which then carry heat flux radially outwards. Since the graphene monolayer is encapsulated by hBN, we assume that the heat flux is generated in the entire graphene/hBN stack rather than just the graphene monolayer. We assume that the phonons propagate ballistically in the individual layers, and also they make transitions between different layers due to scattering at interfaces and disorder, resulting in momentum relaxation at significantly shorter length scales than those for energy relaxation. The heat flux due to this phonon flow can thus be modelled by a 2D heat conduction equation $\vec{j}=-\kappa\nabla T$, where $\vec{j}$ is the radial heat flux and $\kappa$ is the 2D conductivity of the graphene/hBN system. Substituting the continuity equation for the heat flux $\nabla\cdot \vec{j}=P(\mu)\delta(\vec{r})$, with the defect taken to be at the origin, into the heat equation, we obtain
\begin{equation}
\nabla^2 \delta T(\vec{r}) = -\frac{P(\mu)}{\kappa}\delta(\vec{r}) .
\end{equation}
The local temperature change of the graphene/hBN system is 
\begin{equation}
\delta T(r) = \frac{P(\mu)}{2\pi\kappa}\log\frac{L}{r} \,,
\end{equation}
where $L$ is the distance to the heat reservoir for which $\delta T=0$, which is of order the distance to the system edge, and $a$ is the thickness of the graphene/hBN stack. The thermal conductivity $\kappa$ can be written as $\kappa=\frac{1}{2}sc_p' l_{\rm mfp}$, where $c_p'$ is the specific heat capacity of the graphene/hBN system, and $l_{\rm mfp}$ is the momentum relaxation mean free path of the phonons. If we denote $c_p=9\zeta(3)k_B^3 T_p^2/(\pi s^2)$ as the specific heat capacity of the monolayer graphene, we can define the dimensionless ratio $Z\equiv c_p'/c_p$ to account for the contribution from the hBN layers. In the limit where the phonon modes of the different atomic layers are decoupled, $Z$ is equal to the number of atomic layers. A typical value of $r$ would be of order $l_{\rm mfp}$, and we assume $l_{\rm mfp}\sim 2\pi\hbar s/k_BT_p$ to be of the scale of the phonon thermal wavelength. With these assumptions, we obtain
\begin{equation}
\delta T_p = \frac{\hbar}{18\pi \zeta(3)Z} \frac{P(\mu)}{k_B^2 T_p} \log\frac{L}{l_{\rm mfp}} \;\;.
\label{eqn:localT}
\end{equation}

The experimental curve from \cite{halbertal_2017} is plotted in Fig. \ref{fig:PepPartA} for comparison, with the temperature axis shown on the right. We find that the data is consistent with the theoretical prediction for $\epsilon_{\text{LS}}\approx -22$ meV, if we choose $Z\approx 151$, for $L\approx 1 {\rm \mu m}$. Since the thickness of the graphene/hBN system is about 50nm, which corresponds to about 150 atomic layers, this value of $Z$ is physically reasonable. We also note that $\epsilon_{\text{LS}}\approx -22$ meV agrees with the energy of a defect formed by a hydrogen adatom [18]. This highlights the use of resonance cooling as a way to identify the nature of a defect.

\section{Conclusions}

In conclusion, resonant scattering by atomic defects in graphene opens up a new, resonant pathway for electron-lattice cooling. The underlying physics here is phonon emission by electrons trapped on localized states at the defects. This pathway is distinct from those considered before, involving an enhancement of phase space in phonon emission by a free electron in proximity to the defects. The sharp peak in the cooling power $P(\mu)$ near $\epsilon_{\text{LS}}$ enables switching of electron cooling ON and OFF through precise tuning of Fermi energy. In direct analogy with how the Purcell effect is used to control photon emission in optics, resonantly enhanced phonon emission, occurring around the localized defects, can be used to control cooling. One can envisage the design of specific cooling pathways %. This presents 
and, through defect engineering, developing new approaches to control heat flow in nano-systems.
%, and our work shows that graphene is a potential material in this exciting field of application.}

\emph{Note added}. It has come to our attention during the review process that Ref. [27] has come to similar conclusions.
\begin{acknowledgments}
J.F.K. would like to acknowledge financial support from the Singapore A*STAR NSS program. E.Z. acknowledges the support by the Minerva Foundation with funding from the Federal German Ministry of Education and Research. L.S.L. and E.Z. acknowledge the support of the MIT-Israel Seed Fund.
\end{acknowledgments}


\end{document}